\documentclass[10pt,letterpaper]{article}
\usepackage[top=0.85in,left=2.75in,footskip=0.75in]{geometry}
\usepackage{subfig}
\usepackage{amsmath,amssymb}
\usepackage{changepage}
\usepackage[utf8x]{inputenc}
\usepackage{textcomp,marvosym}
\usepackage{cite}
\usepackage{nameref}
\usepackage{hyperref}
\usepackage{booktabs}
\usepackage{bm}

\usepackage{microtype}
\DisableLigatures[f]{encoding = *, family = * }

\usepackage[table]{xcolor}

\usepackage{array}

\newcolumntype{+}{!{\vrule width 2pt}}

\newlength\savedwidth

\raggedright
\usepackage[aboveskip=1pt,labelfont=bf,labelsep=period,justification=raggedright,singlelinecheck=off]{caption}

\makeatletter
\renewcommand{\@biblabel}[1]{\quad#1.}
\makeatother

\usepackage{lastpage,fancyhdr,graphicx}
\usepackage{epstopdf}
\fancyheadoffset[L]{2.25in}
\fancyfootoffset[L]{2.25in}
\usepackage{amsmath,xcolor,colortbl,comment}

\begin{document}
\vspace*{0.2in}

\begin{flushleft}
{\Large
\textbf\newline{Comparative Analysis of Technological Fitness and Coherence at different geographical scales} 
}
\newline
\\
Matteo Straccamore\textsuperscript{1,2,*},
Matteo Bruno\textsuperscript{2},
Andrea Tacchella\textsuperscript{1}
\\
\bigskip
\textbf{1} Centro Ricerche Enrico Fermi, Piazza del Viminale, 1, 00184 Rome, Italy
\\
\textbf{2} Sony CSL - Rome, Joint Initiative CREF-SONY, Piazza del Viminale, 1, 00184 Rome, Italy
\bigskip

* matteo.straccamore@cref.it

\end{flushleft}

\section*{Abstract}
Debates over the trade-offs between specialization and diversification have long intrigued scholars and policymakers. Specialization can amplify an economy by concentrating on core strengths, while diversification reduces vulnerability by distributing investments across multiple sectors. In this paper, we use patent data and the framework of Economic Complexity to investigate how the degree of technological specialization and diversification affects economic development at different scales: metropolitan areas, regions and countries.
We examine two Economic Complexity indicators. Technological Fitness assesses an economic player's ability to diversify and generate sophisticated technologies, while Technological Coherence quantifies the degree of specialization by measuring the similarity among technologies within an economic player's portfolio.
Our results indicate that a high degree of Technological Coherence is associated with increased economic growth only at the metropolitan area level, while its impact turns negative at larger scales. In contrast, Technological Fitness shows a U-shaped relationship with a positive effect in metropolitan areas, a negative influence at the regional level, and again a positive effect at the national level. These findings underscore the complex interplay between technological specialization and diversification across geographical scales.
Understanding these distinctions can inform policymakers and stakeholders in developing tailored strategies for technological advancement and economic growth.


\section{Introduction}
Economic Complexity (EC) is a concept that has gained significant attention in the field of economics in recent years \cite{hausmann2007you, hidalgo2009building, tacchella2012new, hausmann2014atlas, caldarelli2012network}. It refers to the idea that the productive capabilities of countries play a crucial role in determining their economic development and competitiveness in the global market. The concept was initially proposed to argue that traditional measures of economic performance, such as GDP or trade balances, fail to capture the complexity and sophistication of an economy.
At its core, Economic Complexity is based on the notion that the productive structure of an economy is not solely determined by the availability of resources, but also by the diversity and interconnectedness of its industries and technological capabilities \cite{hausmann2014atlas}. Geographical entities such as countries, regions and metropolitan areas that possess a diverse range of industries and a high level of technological know-how are found to have a higher level of Economic Complexity. The importance of EC lies in its potential to explain the long-term economic growth and the development of nations \cite{hidalgo2009building,tacchella2012new}, which brought important institutions such as the World Bank \cite{cristelli2017predictability} and the European Commission \cite{pugliese2020economic, patricia2022china} to adopt the methods and tools of the EC.\\
One of the notable tools in the field of EC is the Fitness and Complexity (EFC) algorithm \cite{tacchella2012new}. This iterative algorithm can be employed to analyze bipartite networks, i.e. systems with two different types of actors interacting between them, and was originally developed to assess a country's export capacity (Fitness) and measure the complexity of the exported product (Complexity). More tools based on networks \cite{hidalgo2007product, zaccaria2014taxonomy} or machine learning \cite{albora2023product, fessina2024identifying, albora2022machine, tacchella2023relatedness} have been developed to leverage the concept of similarity and predict a country's future range of exportable products. The idea underlying these studies is that capabilities possessed by individual countries are reflected in their ability to export products that are similar to each other: when sharing a relatively large number of capabilities products are likely to be exported together \cite{saracco2015innovation}. Fitness is a generalized measure of the diversification of an economic actor. In particular, it weights diversification by the complexity of its economic activities. In turn, complex activities are defined as those only performed by actors with high Fitness. Fitness is considered a proxy for the Economic Complexity of an economic actor, namely its ability to combine a large set of endowments to produce complex outputs. This measure is an extremely effective predictor of economic performance \cite{cristelli2017predictability}.\\
On the other hand, Coherence is a second key measure presented by Economic Complexity. Introduced by Pugliese et al. \cite{pugliese2019coherent}, it quantifies the average distance between the economic activities in which an actor is involved. This distance is measured on a similarity network that captures the relationships between economic activities in terms of their underlying capability requirements \cite{zaccaria2014taxonomy, pugliese2019coherent, straccamore2023urban}. Compared to Fitness, which is more related to the diversification of an economic actor, Coherence is a measure of how much more specialized the actor is in a given economic activity.\\
EC is not solely based on production and also patent data and technology innovation can be used to measure innovation activity. The literature extensively acknowledges the use of patent data for tracking technological advancements~\cite{frietsch2010value, griliches1998patents, leydesdorff2015patents}. Pugliese et al. \cite{pugliese2019unfolding} further demonstrated that technological metrics serve as superior predictors for industrial and scientific output in the forthcoming decades.
In addition, patent data increased accessibility and its geolocalization have transformed patents into a crucial resource for studying technological evolution \cite{youn2015invention, de2019geocoding}. Finally, a key feature of patent documents significant for economic analysis is the inclusion of classification codes, which categorize inventions into specific technological sectors \cite{fall2003automated, falasco2002bases, falasco2002united}. These classification systems provide detailed descriptions of technological fields, facilitating targeted analyses within particular domains.
However, using patents as a proxy for innovation measurement has notable limitations \cite{hall2014choice} such as that only a subset has a significant market value \cite{hall2005market}, or the misleading in aggregate patent statistics to represent economic and inventive activities \cite{pavitt1985patent}. Moreover, patents do not represent all facets of knowledge production within the economy \cite{griliches1998patents}, do not encompass all knowledge generated \cite{arts2013inventions} and do not represent all economic sectors \cite{kogler2015intellectual, lanjouw1996innovation}.\\

The EC approach was used with technology codes on different geographical scales, analyzing firms \cite{pugliese2019coherent, straccamore2022will, arsini2023prediction, albora2024machine}, metropolitan areas \cite{straccamore2023urban, balland2017geography, boschma2015relatedness, kogler2018patent, kogler2013mapping, balland2015technological}, and regions~\cite{oneale2021structure, napolitano2018technology, dettmann2013determinants, tavassoli2014role, colombelli2014emergence, sbardella2021behind, sbardella2018green, sbardella2017economic}.
Despite their wide application, a comparison of the insights provided by EC instruments at different geographical scales is still lacking. This study contributes to the literature on Economic Complexity by comparing for the first time, to the best of our knowledge, the heterogeneous relations between Economic Complexity indicators and growth at different geographical scales.\\
We utilize the EFC algorithm to compare technological outputs at different geographic scales, including metropolitan areas (MAs), regions, and countries (from now on, unless specified, by the term \textit{entity} we will mean without change in meaning, either MAs or regions or countries). We aim to compare the correlation levels between Technological Fitness (F) and Technological Coherence ($\Gamma$) \cite{pugliese2019coherent,straccamore2023urban} with economic growth, measured by the GDP per capita (GDPpc) of these entities. In our results, Fitness and Coherence display a rich and heterogeneous correlation structure, which has, to the best of our knowledge, not been studied extensively in the literature. Here we present a first attempt to assess their joint impact on economic growth, and how this relationship changes with the scale of the entities considered. We find that they play different roles, with different scaling relations.
Previous research \cite{straccamore2023urban} has shown that $\Gamma$ plays a more significant role in predicting GDP per capita growth for MAs. In this paper, we emphasize this finding by examining the importance of different features. Additionally, as we expand our analysis to regions and countries, we observe how $\Gamma$ has a different impact on the economic growth of all three geographical scales.
In particular, we find that Coherence has a positive effect only at the MA level, but its impact turns negative at the regional scale, a finding that becomes even more pronounced at the national level. The results for Fitness are less clear-cut: it exhibits a U-shaped relationship with a positive association with growth at the MA level, a negative correlation with growth at the regional scale, and again a positive correlation at the national level.
By comprehending the distinct roles played by Technological Fitness and Coherence in various contexts, policymakers and stakeholders can develop customized strategies to promote technological progress and regional growth.\\
This paper is organized as follows. In Section \ref{sec:data}, we describe the data sources: patent and technology codes, GDP per capita information and metropolitan areas and region boundaries data. Section \ref{sec:methods} outlines the methodological framework, including the construction of bipartite networks, the implementation of the Fitness and Complexity algorithm, and the computation of Technological Coherence. In Section \ref{sec:results}, we present our empirical results, highlighting the distinct roles of Technological Fitness and Coherence across different geographical scales. Finally, Section \ref{sec:discussion} concludes the paper and summarizing the key contributions of the research.

\section{Data}
\label{sec:data}
\subsection*{Technology Codes}

In this research, we employed the PATSTAT database (available at \url{www.epo.org/searching-for-patents/business/patstat}) as our main source for patent and technology code data. This database contains about 100 million patents registered across 100 global Patent Offices. Each patent is uniquely identified and associated with multiple Cooperative Patent Classification (CPC) codes \cite{montecchi2013searching}. The World International Patent Office (WIPO) uses the CPC system, which is more detailed than the previously used International Patent Classification (IPC) system \cite{fall2003automated}. The CPC provides a hierarchical classification scheme that includes sections, classes, subclasses, and groups, facilitating fine-grained categorization of patents. The first level of CPC codes indicates broad technology categories, such as "Chemistry; Metallurgy" under the code C or "Electricity" under the code H. Subsequent levels further specify the technology, for instance, "Inorganic Chemistry" under C01 or "Organic Chemistry" under C07. Each patent is also assigned a filing date based on its initial submission.

For the geolocation of patents, we used the De Rassenfosse et al. database \cite{de2019geocoding}, which includes data on 18 million patents spanning from 1980 to 2014 and provides precise geographical coordinates for each patent, enabling accurate geolocation. In our analysis, each patent is linked with a unique identifier, a set of CPC codes, and geographic coordinates that identify the corresponding MA, region, and country. The De Rassenfosse et al. database also includes the country ISO code, which aids in constructing the country-technology bipartite graph. Additional details about the importance and features of the De Rassenfosse et al. database are discussed in the Supplementary Information.

\subsection*{GDP per capita}

To retrieve data on the Gross Domestic Product (GDP) per capita for countries, we utilized the World Bank's comprehensive database, World Development Indicators (WDI), accessible at \url{data.worldbank.org}. The WDI provides extensive GDP information for countries worldwide.\\
Regarding the GDP per capita of metropolitan areas (MAs) and regions, we referred to the work of Kummu et al.~\cite{kummu2018gridded}. Their study aimed to create high-resolution global datasets for GDP and Human Development Index (HDI) spanning the period from 1990 to 2015. By integrating various data sources such as national accounts, satellite imagery, and geospatial datasets, the authors developed gridded datasets that accurately portray the spatial distribution of GDP and HDI indicators at a fine resolution. Statistical techniques and modelling approaches were employed to estimate GDP and HDI values for areas lacking direct measurements. The Kummu et al. database is all in 2011 international U.S. dollars at purchasing power parity.

\subsection*{Metropolitan areas and regions boundaries}
In order to calculate the Gross Domestic Product per capita (GDPpc) and link patents to specific metropolitan areas (MAs) and regions, it was necessary to have access to the geographical boundaries of these entities.
Regarding MAs, we downloaded the boundaries from the Global Human Settlement Layer (GHSL)~\cite{Schiavina2019GHS}. The GHSL provides global spatial information on human settlements, including the boundaries of metropolitan areas. The GHSL dataset incorporates satellite imagery, census data, and other sources to map urban areas and classify them into different settlement types. It offers valuable information for various research and planning applications, including urban studies, infrastructure development, and environmental analysis.\\
Regarding the boundaries of the regions, we downloaded them from the Global Administrative Areas (GADM)~\cite{areas2012gadm} website. GADM provides administrative boundary data for regions, countries, and other administrative divisions worldwide. It is a reliable and widely used resource for accessing shapefiles or spatial data of administrative boundaries. The GADM database contains detailed and up-to-date boundary information for regions across the globe. It is a valuable tool for various applications, including geographic analysis, research, and mapping. The database offers different administrative levels, allowing users to access region-level boundaries as well as higher or lower administrative divisions depending on their needs. We selected the administrative level 1 of the GADM database, the first subdivision below the country level.\\
After obtaining the boundaries, we are able to associate the patents and calculate the GDPpc for each of them.  Starting from the patents' geographical coordinates, we can select all those that fall within a specific boundary and, consequently, which technology codes. To compute the GDP per capita of each MA and region, we consider the GDP grid in one year, we compute the GPD per capita of a MA or region as the average of all the grid points within its boundaries.

\section{Methods}
\label{sec:methods}
\subsection*{Bipartite networks construction}
After collecting the data, we construct bipartite networks linking metropolitan areas (MAs), regions, or countries with technology codes. These networks are represented by bi-adjacent rectangular matrices, $\textbf{M}^y$, where each element $M_{e,t}^y$ is set to $1$ or $0$. This indicates whether the presence of a technology code $t$ in patents filed by entity $e$ is statistically significant for the year $y$.

To build the $\textbf{M}^y$ matrices, we employed an innovative methodology involving statistical tests to validate the number of technologies present in an entity's patents. We start with two preliminary matrices: $\textbf{V}^y$, linking entities $e$ to their patents $p$, and $\textbf{B}^y$, linking patents to their corresponding technologies $t$. In these matrices, $V_{e,p} = 1$ (or $0$) indicates whether patent $p$ is made by entity $e$, and $B_{p,t} = 1$ (or $0$) if patent $p$ includes technology code $t$. Given all patents belonging to an entity $e$, we count the number of observations of each technology, i.e. the number of times technologies $t$ fall into the entity $e$ for all $p$ associated with $e$. We validate these counts by comparing them against an ensemble of $1000$ matrices generated using the Bipartite Configuration Model (BiCM) \cite{saracco2017inferring, saracco2015randomizing}, thus obtaining a matrix of p-values, $\textbf{PVAL}^y$. We compute the BiCM by using the \textit{NEMtropy} Python package (\url{github.com/nicoloval/NEMtropy}) \cite{vallarano2021fast}. Each element, $\text{PVAL}_{e,t}^y$, represents the p-value associated with the link between entity $e$ and technology $t$ for year $y$. In Supplementary Information, the construction process for $\textbf{PVAL}^y$ matrices is dissed in more detail.

Initially, there were $8641$ MAs listed in the Global Human Settlement Layer~\cite{Schiavina2019GHS}, covering the entire globe. To minimize statistical fluctuations, we keep the top 80\% of entities by number of patents. In particular, we computed the mean diversification corresponding to the top 80\% entities for each year, and selected the diversification threshold as the mean during the years for each entity. Moreover, we excluded technologies appearing in fewer than $100$ patents. This adjustment reduced the number of patent-producing MAs to $1847$. Additionally, for some of the MAs and regions, the computation of the GDPpc is not possible due to their small dimensions. In particular, this appears for $277$ MAs and $8$ regions. Consequently, the network finally comprises $1570$ MAs, $738$ regions, and $46$ countries, linked to $621$ distinct technology codes.

To account for the volatility of year-to-year data, we consider a rolling window of five years for constructing both $\textbf{V}^y$ and $\textbf{B}^y$. Thus, in this study, $\textbf{PVAL}^y$ and $\textbf{M}^y$ refer to five-year periods, ranging from $1980-1984$ to $2010-2014$. The dataset comprises $31$ in such five-year window matrices.

Lastly, we binarize $\textbf{PVAL}^y$ using a p-value threshold of $0.05$ as follows:
\begin{equation*}
    M^y_{e,t} = 
    \begin{cases}
        1\ \text{if}\ \text{PVAL}^y_{e,t} \le 0.05\\
        0\ \text{if}\ \text{PVAL}^y_{e,t} > 0.05.
    \end{cases}
\end{equation*}

\subsection*{Fitness and Complexity algorithms}
The Fitness and Complexity (FC) framework \cite{tacchella2012new}, introduced in 2012, provides a method for quantifying the competitiveness (Fitness) of a country's economy. In this study, we apply the FC framework to quantify the Technological Fitness of metropolitan areas (MAs), regions, or countries, considering patent data exclusively. The iterative process for determining these quantities is as follows:
\begin{equation}
\begin{cases}
\tilde{F}_e^{n+1} = \sum_t{M_{et}Q^{n}_t}\\
\tilde{Q}_t^{n+1} = \left(\sum_e{M_{et}(F^{n}_e)^{-1}}\right)^{-1}
\end{cases}
\label{eq:FCalg}
\end{equation}
Here, in each iteration step $n$, the quantities are normalized:
\begin{equation}
\begin{cases}
F^n_e = \frac{\tilde{F}_e^n}{\left<\tilde{F}\right>_e}\\
Q_t^n = \frac{\tilde{Q}_t^n}{\left<\tilde{Q}\right>_t}
\end{cases}
\end{equation}
The initial conditions are set as $Q_t^{(0)} = 1$ $\forall t$, and $F_e^{(0)} = 1$ $\forall e$. Extensive studies have been conducted on the convergence of this algorithm \cite{pugliese2016convergence}. In our case, we compute $F^y_e$ and $Q^y_t$ for each 5-year window $y$ using the bi-adjacency matrices $M^y_{et}$. The iteration process is stopped when there is no further change in the Fitness ranking of entities.
The rationale behind this approach is as follows: a technology developed in an already advanced entity provides limited information about the complexity of the technology itself since advanced entities produce a large proportion of technologies. In contrast, a technology exported by an underdeveloped entity is likely to possess a lower level of sophistication. Hence, an entity's technological competitiveness can be measured based on the complexity of its technologies. However, a different approach is required to assess technology quality. Fitness $F_e$ is proportional to the sum of technologies, weighted by their complexity $Q_t$. Intuitively, the Complexity of a technology is inversely proportional to the number of entities that have implemented it. If an entity has a high Fitness value, it should contribute less to limiting the complexity of the technology, while entities with low Fitness values should strongly influence $Q_t$.

\begin{figure*}
    \centering
    \includegraphics[scale=.75]{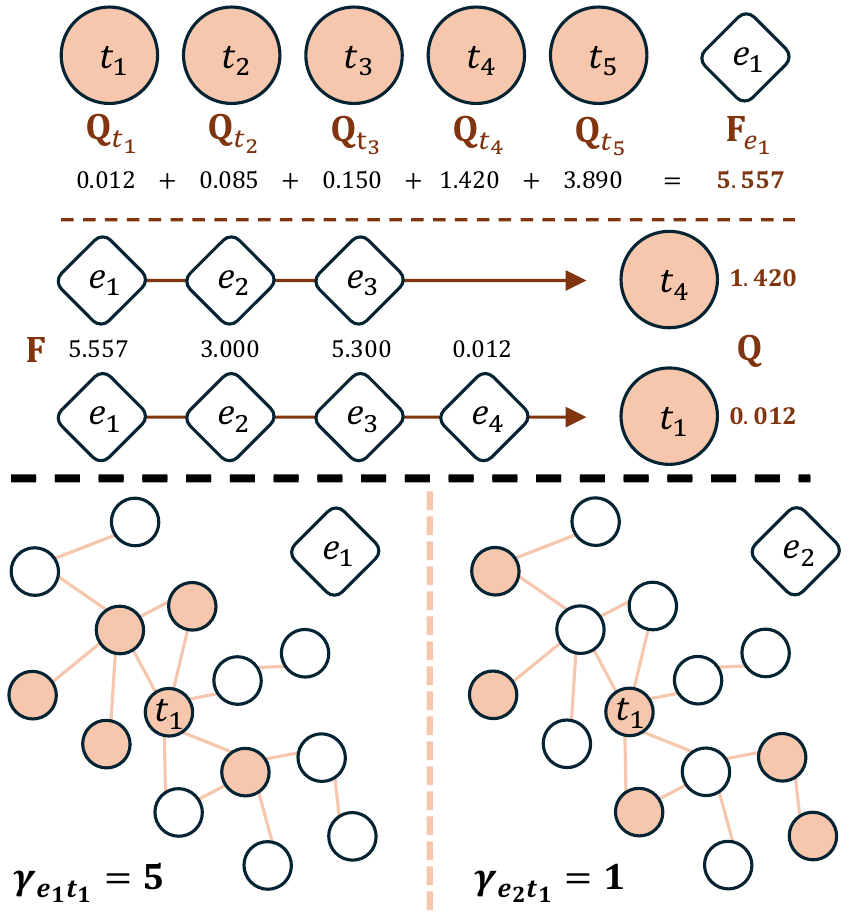}
\caption{\textbf{Technological Fitness and Coherence explanation.} \textbf{Schematic representation of the workings of the Fitness and Complexity algorithm and Coherence diversification.} In the top panel, we show the computation of the Fitness and Complexity algorithm. The Fitness of the entity $e_1$ is the technological Complexity sum of all the technology done by $e_1$. Regarding the Complexity, for $t_4$ it is high because entities do it with high Fitness. Instead, $t_1$ has a low Complexity because is done by $e_4$ which has a low Fitness. In other words, we can see $e_1$, $e_2$ and $e_3$ as technologically advanced entities, so if $t_4$ is made only by these it means that it is a technology that is, indeed, complex. Unlike the previous ones, $e_4$ does not have the same technological development, and since this one can produce $t_1$ together with the other entities it implies that $t_1$ has low Complexity. At the bottom, we explain the idea of Coherence diversification. Both figures offer a simplified representation of the network of technologies $\textbf{S}$. Each node is a technology and the links quantify if technologies are connected, i.e., if they are similar and if the capabilities required to develop both are approximately the same. The coloured nodes are the technologies made by entity $e_1$ and $e_2$, respectively. Entity $e_1$ has a high Coherence value concerning $t_1$ since its technology portfolio has very closely related elements. In contrast, $e_2$ has a portfolio with disconnected technologies and consequently has a low Coherence value.}
\label{fig:F_C_expl}
\end{figure*}

\subsection*{Coherent Diversification}
Previous studies have highlighted the significance of coherence in production and innovation diversification as a key driver of productivity \cite{quatraro2010knowledge, kalapouti2015intra}. Thus, to gain a better understanding of the performance of our entities based on their technology portfolio, we examine their coherent diversification \cite{pugliese2019coherent}. The central question is whether the accumulation of knowledge and capabilities associated with a coherent set of technologies leads metropolitan areas (MAs), regions, or countries to experience greater benefits in terms of GDP per capita. Coherent diversification is defined as the coherence of a technology field $t$ concerning the technology basket of entity $e$:
\begin{equation}
\gamma_{et} = \sum_{t' \neq t} {S^y_{tt'}M^y_{et'}},
\label{eq:coer1}
\end{equation}
where $M$ is the bipartite adjacency matrix between MAs, regions, countries, and technologies. $\textbf{S}^y$ represents the similarities matrix between pairs of technologies and is computed with the same statistical method used to build $\textbf{M}$ with the same p-value threshold $0.05$. However, for the $\textbf{S}^y$ computation, we used only the patent-technology matrices for each 5-year window. To have a clear interpretation, we statistically validated how many times two technologies $t$ and $t'$ appear in different patents.\\

For each technological field $t$ and each entity $e$, we count the number of technologies $t'$ adopted by $e$ that are connected to $t$ using the $S^y_{tt'}$. If the technological portfolio of entity $e$ consists of numerous strongly connected technologies surrounding $t$, then $t$ will exhibit high coherence to $e$, resulting in a high value of $\gamma_{et}$. Conversely, if $t$ belongs to a portion of the technology network that is distant from the patenting activity of entity $e$, $\gamma_{et}$ will be low. It is important to note that $\bm{\gamma}$ has the same dimensions as $\textbf{M}$, with its elements quantifying the coherence of a technology $t$ to the technology basket of entity $e$.
Finally, we can calculate the Technological Coherence of entity $e$ \cite{pugliese2019coherent} as follows:
\begin{equation}
\Gamma_e = \frac{\sum_t{M_{et}{\gamma_{et}}}}{d_e^2},
\label{eq:coer2}
\end{equation}
where $d_e = \sum_t{M_{et}}$ represents the diversification of entity $e$. The Technological Coherence ($\Gamma_e$) computes the average coherence $\gamma$ of the technologies in which entity $e$ is engaged in patenting activities. Compared to the original definition \cite{pugliese2019coherent}, we have divided the following quantity twice by $d_e$. The reason is related to the square dependence of $d_e$ on diversification. One contribution comes from $\textbf{M}$ in the definition of $\bm{\gamma}$, the second from $\textbf{M}$ in $\Gamma$. By dividing by $d_e^2$, we eliminate this dependence and are better able to measure the coherence of an entity $e$.

In Fig. \ref{fig:F_C_expl}, we present a pictorial representation of how Technological Fitness and Coherence work. In the upper panel, the calculation of the Fitness and Complexity algorithm is presented. The Fitness of entity $e_1$ represents the cumulative technological Complexity of all technologies developed by $e_1$. As for Complexity, technology $t_4$ is considered highly complex because it is pursued by entities with high Fitness. Conversely, technology $t_1$ is deemed to have low complexity since it is undertaken by $e_4$, which possesses low Fitness. Thus, entities $e_1$, $e_2$, and $e_3$ can be viewed as technologically sophisticated, suggesting that if only they are involved in producing $t_4$, it underscores the complexity of the technology. On the other hand, $e_4$ lacks comparable technological advancement. Therefore, its ability to produce $t_1$ alongside other entities indicates that $t_1$ is less complex. The lower panel elucidates the concept of Technological Coherence. Both diagrams offer a streamlined depiction of the technological network $\textbf{S}$. Each node corresponds to a technology and the connections show whether there is significant similarity between the technologies, reflecting whether or not their development requires similar capabilities. The technologies attributed to entities $e_1$ and $e_2$ are shown as coloured nodes. Entity $e_1$ exhibits a high Coherence value for $t_1$ due to the closely related technologies within its portfolio. In contrast, the portfolio of $e_2$ features unconnected technologies, resulting in a lower Coherence value.

\begin{figure*}
\centering
\subfloat[]
   {\includegraphics[width=.65\textwidth]{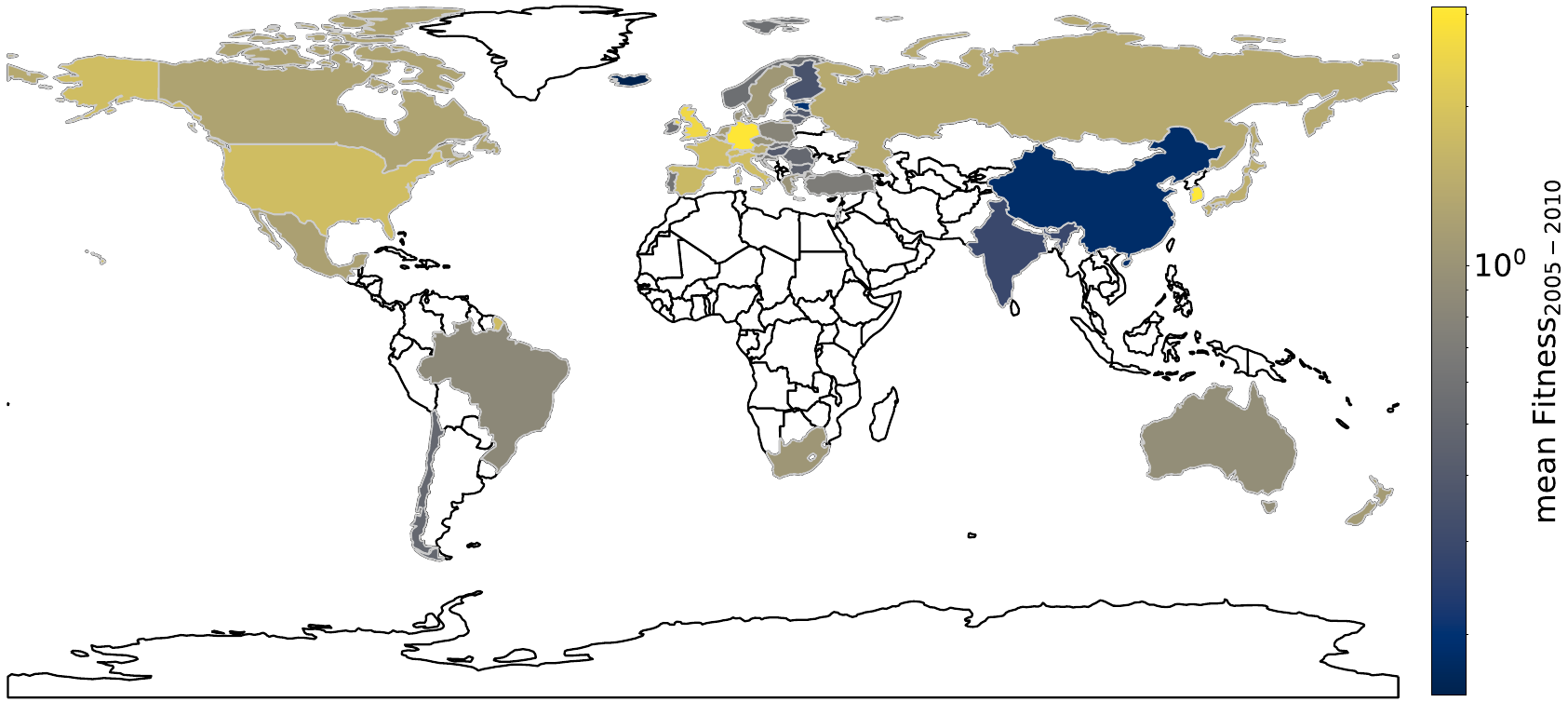}} \quad
\subfloat[]
   {\includegraphics[width=.65\textwidth]{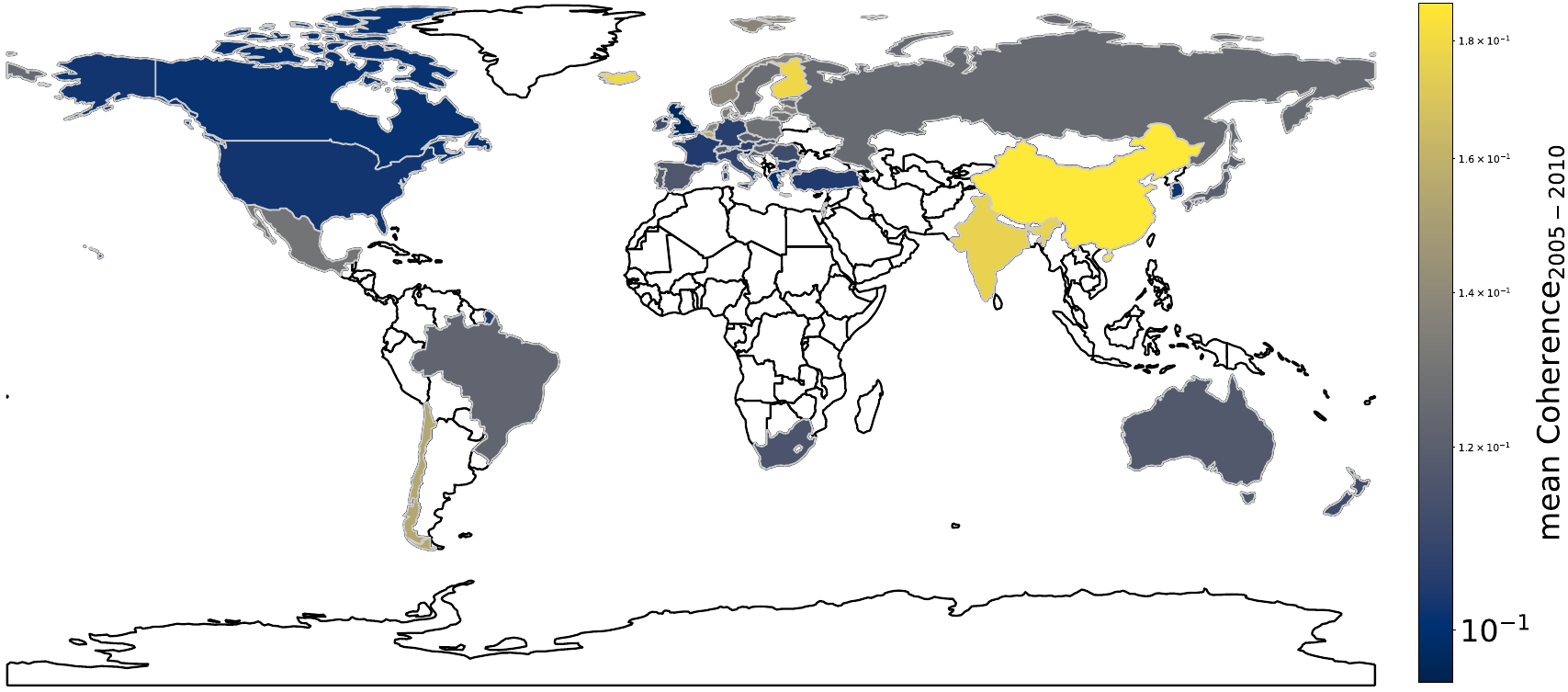}} \\
\subfloat[]
   {\includegraphics[width=.45\textwidth]{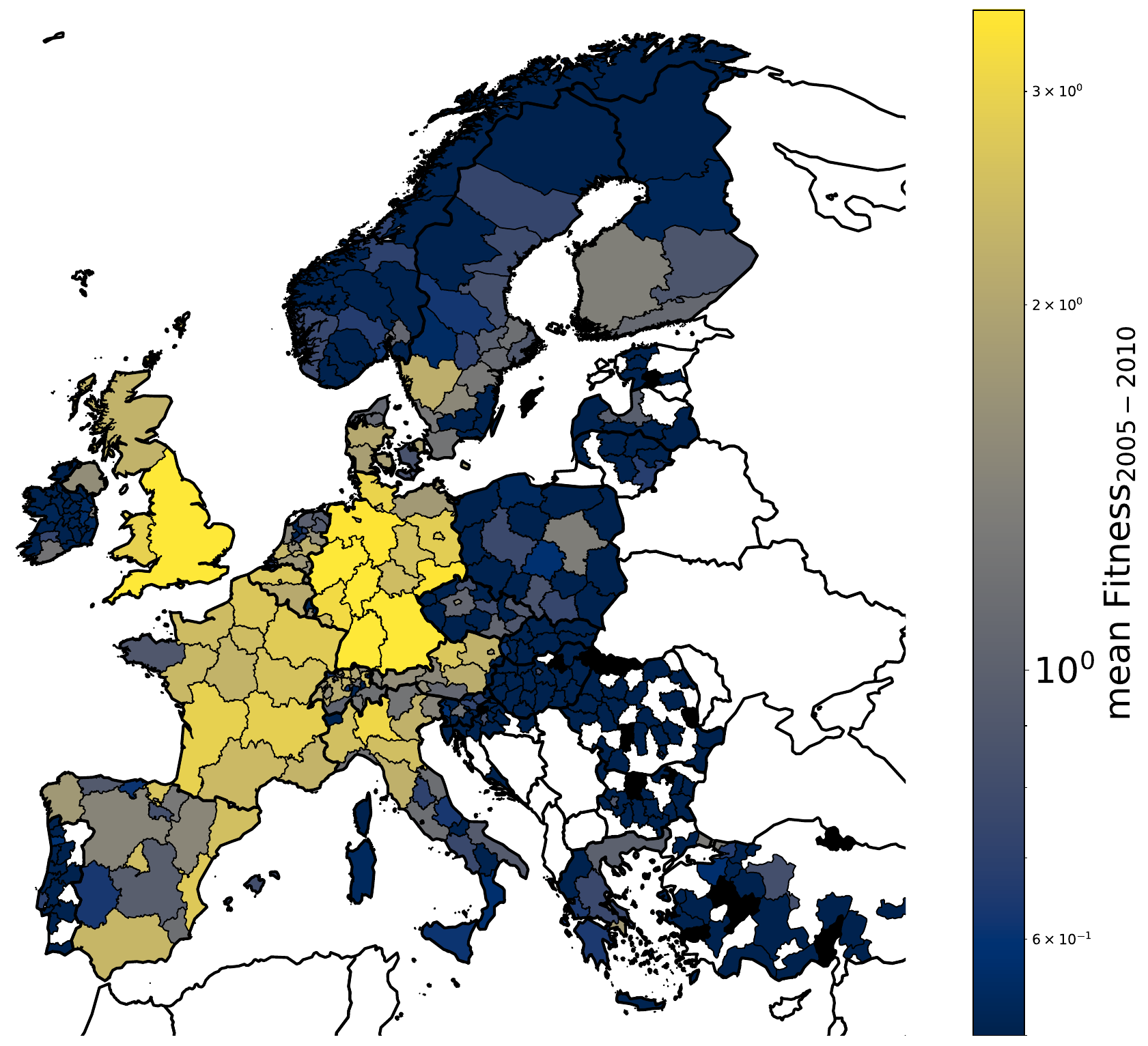}} \quad
\subfloat[]
   {\includegraphics[width=.45\textwidth]{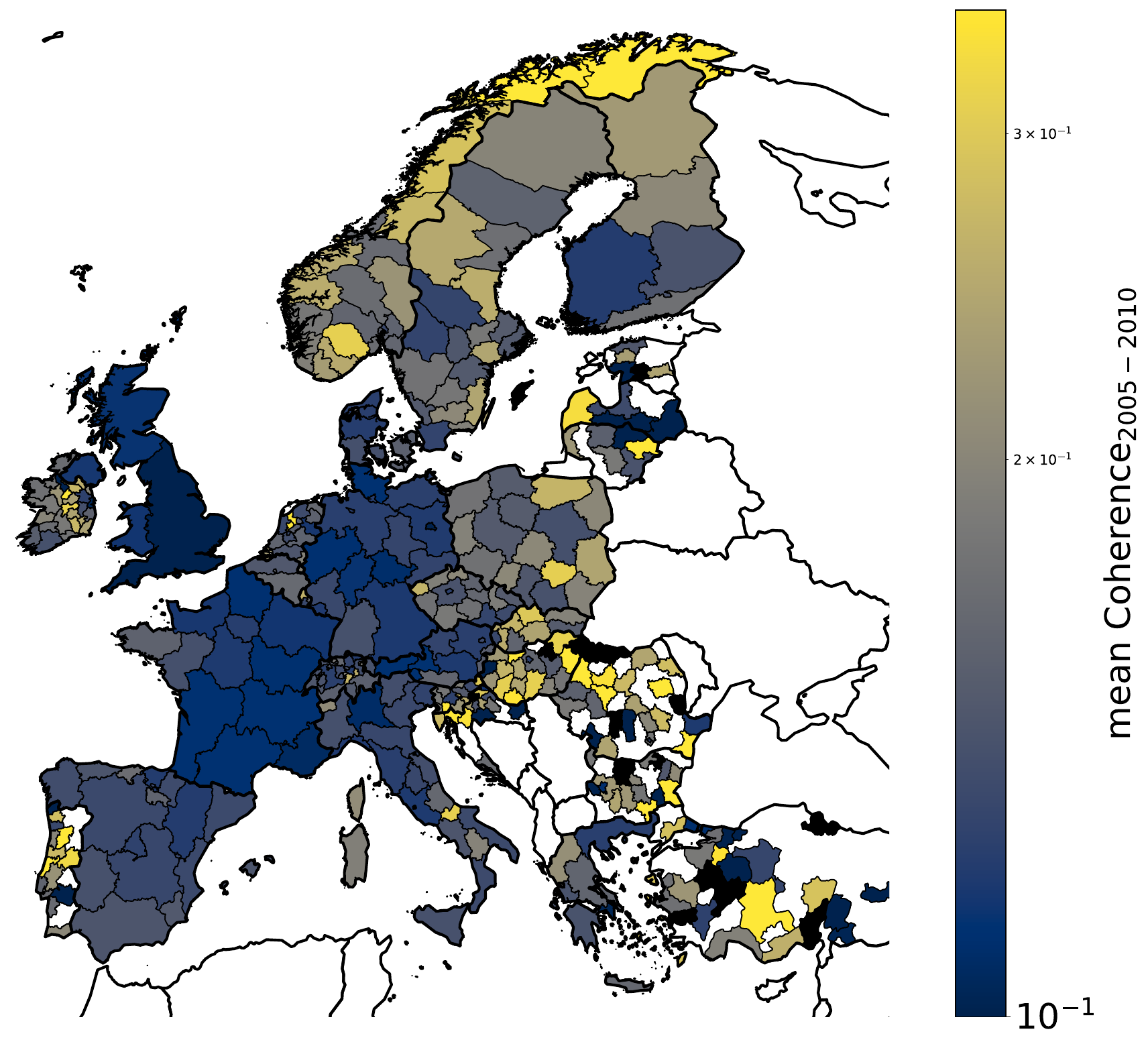}}
\caption{\textbf{Mean Technological Fitness and Coherence in the world and European regions.} We compute the mean Technological Fitness and Coherence from $y_0 = 2005$ to $y_1 = 2010$ considering the country case (\textbf{(a)} for Fitness and \textbf{(b)} for Coherence) and the European regions case (\textbf{(c)} For Fitness and \textbf{(d)} for Coherence). We highlight how the two measures are not directly related. For example, China has high Coherence and low Fitness. In contrast, at the European level, regions in the West have high Fitness while those in the East have high Coherence.}
\label{fig:F_C_W_reg}
\end{figure*}

In Fig. \ref{fig:F_C_W_reg}, we plot the World and Europe maps colouring each country in our database or European region according to the mean technology's Fitness and Coherence from $y_0 = 2005$ to $y_1 = 2010$ values.
We can note that the two measures are not directly related. For example, China has high Coherence and low Fitness. In contrast, at the European level, regions in the West have high Fitness while those in the East have high Coherence. As an example, we report some of the domains in which China and Eastern European regions demonstrate high Coherence. In our investigation, we observed significant Coherence in China's technological production, particularly in the F21 (Lighting) and H04 (Electric Communication Technique) categories. Additionally, the regional analysis highlighted several areas of noteworthy Coherence. For instance, the Karlovarský region in the Czech Republic demonstrated high coherence in the production of technologies within the E01 (Construction of Roads, Railways, or Bridges) category, as well as in the Physics and H04 categories. Similarly, Gävleborg in Sweden exhibited coherence across diverse sectors, including Personal or Domestic Articles, Health; Life-Saving; Amusement, Shaping, Metallurgy, and Earth or Rock Drilling; Mining. In Finland, the Oulu region showed pronounced coherence in H03 (Basic Electronic Circuitry) and H04. Notably, it also displayed strong production capabilities in A01 (Agriculture; Forestry; Animal Husbandry; Hunting; Trapping; Fishing), A23 (Foods or Foodstuffs), A61 (Medical or Veterinary Science; Hygiene), along with Chemistry and Physics categories.


\section{Results}
\label{sec:results}
In this section, we present our findings on the impact of technological Fitness and Coherence on GDP growth at different scales. We quantify economic growth as the Compound annual growth rate (CAGR) of GDPpc from $y_1$ to $y_1 + \delta$. CAGR is defined as:
\begin{equation}
    \mathrm{CAGR}(y_0,y_1) = \left(\frac{\mathrm{GDPpc}(\mathrm{y_1})}{\mathrm{GDPpc}(\mathrm{y_0})}\right)^{\frac{1}{y_{1}-y_{0}}}-1
\end{equation}

In this Section, we present the results following this scheme:
\begin{itemize}
    \item \textbf{Initial Overview:} We begin by presenting a pairwise correlation analysis between Fitness, Coherence and CAGR. This analysis provides a qualitative description of these interactions and demonstrate non-trivial and heterogeneous correlation patterns. These findings are presented in Fig. \ref{fig:F_C_CAGR}.
    \item \textbf{Feature Importance Analysis:} In this second step we consider the impact of Fitness and Coherence in the prediction of future CAGR, in a non-linear multivariate setting, with multiple controls. The results of this analysis are presented in Fig. \ref{fig:RF_FI} and \ref{fig:SHAP}
\end{itemize}

\subsection{Initial Overview}
\begin{figure}
    \centering
    \includegraphics[scale=.70]{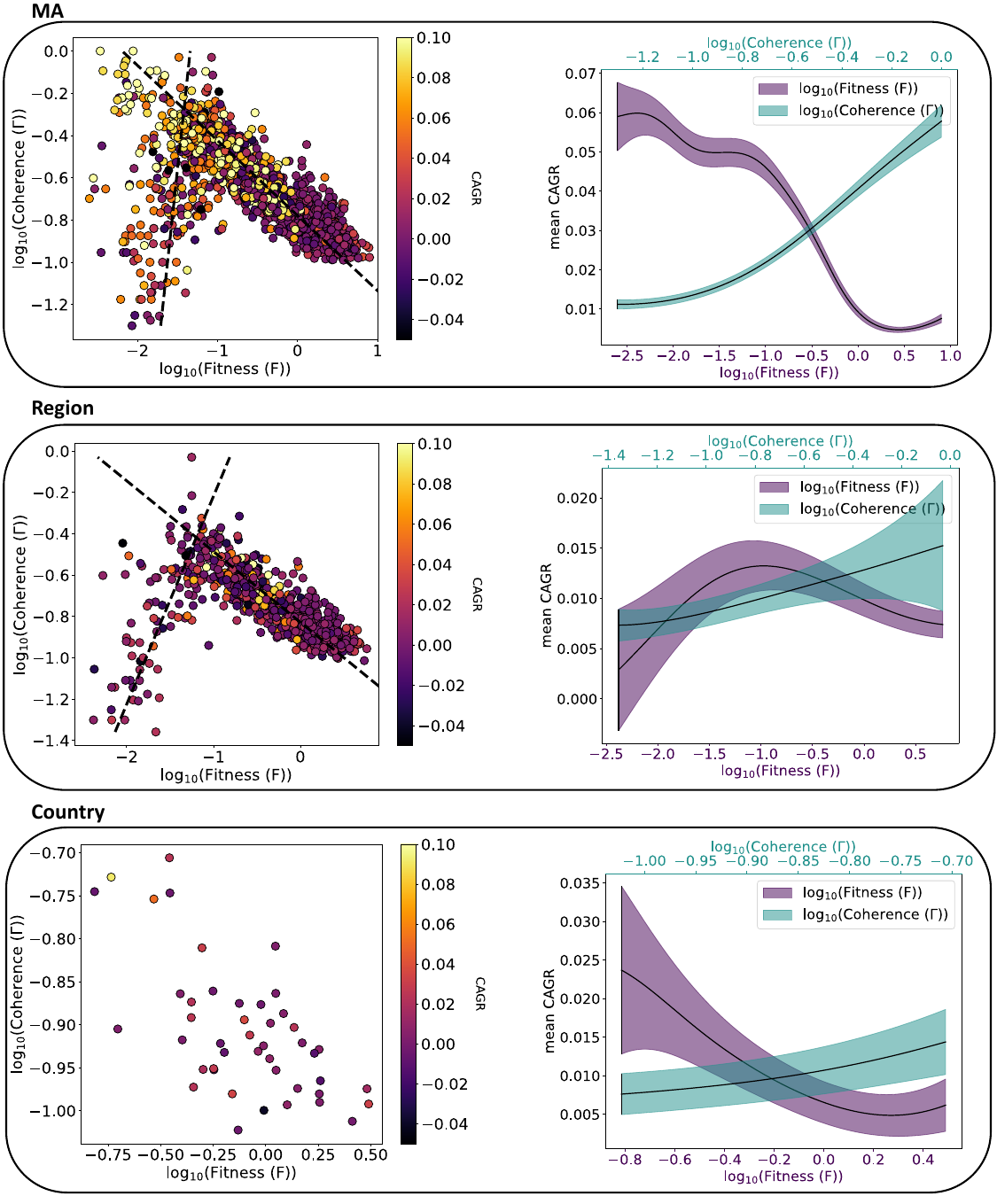}
\caption{\textbf{Technological Fitness VS Technological Coherence to evaluate GDPpc growth.} We compute the mean Technology Fitness and Coherence for $y_0 = 2005$ and $y_1 = 2010$. The left panels are the scatter plot of the previous values and the colour bar is the CAGR($y_0$,$y_1$). In the right panels, we calculate the means and the standard deviation of the mean of the CAGR($y_0$,$y_1$) weighting the points with a Gaussian Kernel centred in $200$ parts between the minimum and maximum value of both Fitness and Coherence of the points in the left panels. In all three cases, we observe the same trend: the Fitness curve tends to decrease, less evident in region case where it has a concave behavior. The Coherence has a convex one, more evident in the case of MAs and less so in the case of countries. In addition in regions and MAs, we highlight two distinctive trends that could suggest two types of economic development.}
\label{fig:F_C_CAGR}
\end{figure}

As an initial step, we want to highlight the relation between Fitness, Coherence and CAGR. To do this, we decided to present a snapshot of these relations in the period 2005-2010. In Fig. \ref{fig:F_C_CAGR}, we plot the mean Fitness and Coherence for $y_0 = 2005$ and $y_1 = 2010$ as a function of CAGR(2005,2010). Panels \textbf{a}, \textbf{b}, \textbf{c} are respectively for MA, region, and country. 
In all three cases, we observe a similar trend: the Fitness curve tends to decrease, less evident in region case where it has a concave behavior. The Coherence has a convex one, more evident in the case of MAs and less for countries. Finally in regions and MAs, we decided to highlight two distinctive trends that could suggest two types of economic development.

\subsection{Feature importance with SHAP analysis}
To determine the impact of Technological Fitness ($F$) and Technological Coherence ($\Gamma$), we treat our problem as a regression problem. Our independent variables, or features, are given by the average of Technological Fitness and Technological Coherence over a range of years from $y_0$ to $y_1 = y_0 + \delta$, $\overline{F}(y_0,y_1)$ and $\overline{\Gamma}(y_0,y_1)$; the dependence variable, on the other hand, is represented by the Compound annual growth rate (CAGR) of GDPpc from $y_1$ to $y_1 + \delta$. We employ a Random Forest Regression model and conduct a feature importance analysis. This allows us to quantify their respective roles in the analysis. RandomForest algorithm \cite{breiman2001random} is a tree-based machine learning algorithm used to capture the non-linear links between variables better. The field of Economic Complexity literature has demonstrated its utility in making accurate predictions and effectively capturing the nonlinear relationships between variables used in Economic Complexity \cite{straccamore2022will, tacchella2023relatedness, albora2022machine, albora2023product, fessina2024identifying}.\\
Given the high correlation between Fitness and diversification, and the high dependence on the starting GDPpc, we also include these quantities in the analysis to control both $F$ and $\Gamma$. In addition for this purpose, we control taking into account geographical fixed effects (MA$_{\textbf{fe}}$, region$_{\textbf{fe}}$ and country$_{\textbf{fe}}$) and time fixed effect $y_0$.
\begin{figure}
    \centering
    \includegraphics[scale=.6]{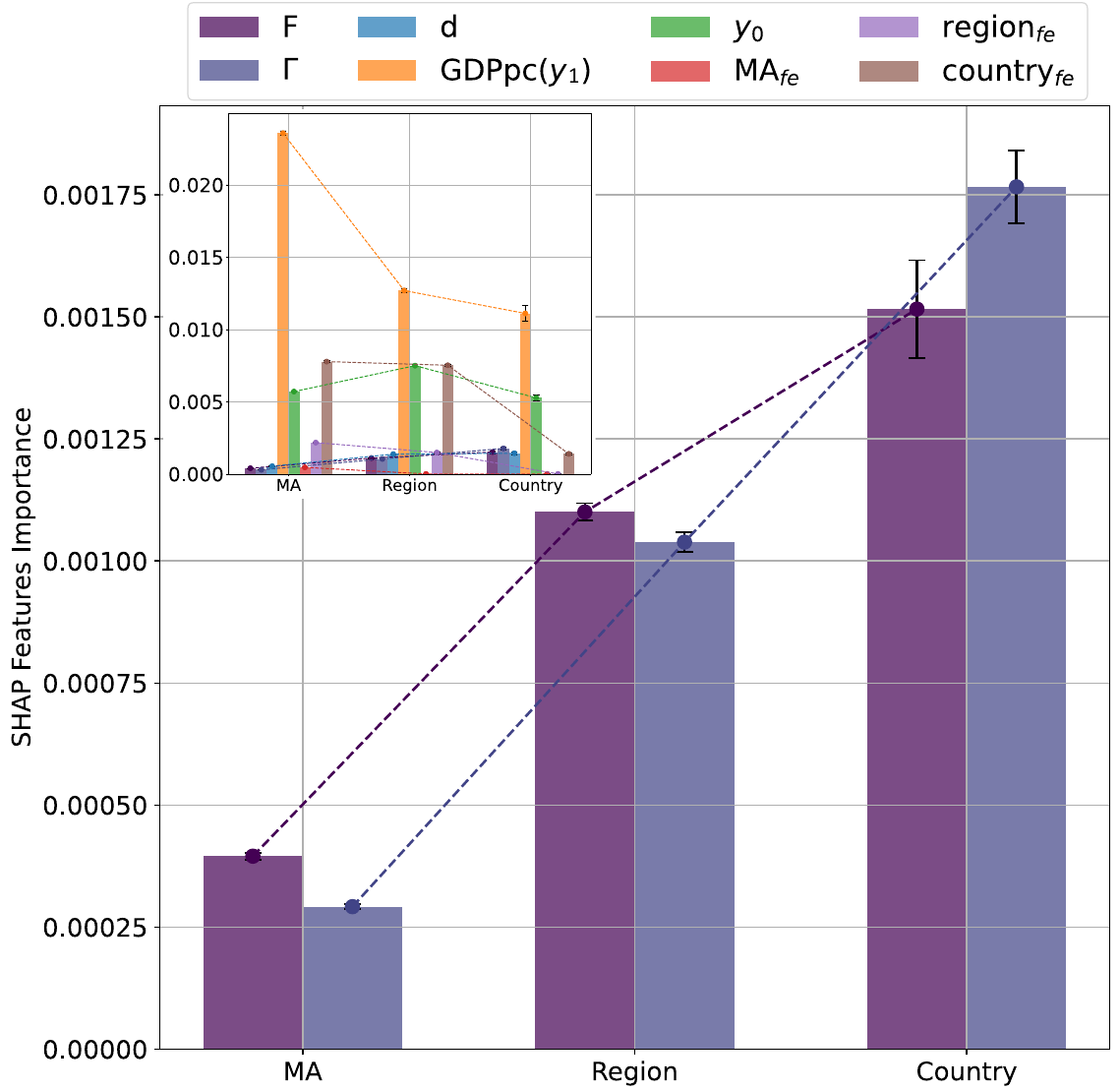}
\caption{\textbf{SHAP values feature importance.}
The feature importance of Fitness, Coherence, Diversification, starting GDPpc, and geographic and time information is depicted by each pair of bars. We notice how the feature GDPpc($y_1$) is the most significant followed by year and country geographic information. We incorporated them solely to control the behaviours of Fitness and Coherence. A focused comparison of these two metrics is illustrated in the main Figure. We find that Fitness and Coherence relative importance switch from MA to the country in explaining the trends in CAGR. However, we do not yet know whether this importance translates into a positive or negative effect.}
\label{fig:RF_FI}
\end{figure}
First, we use the "RandomForestRegressor" from the "sklearn" python library \cite{pedregosa2011scikit} to train our regression model:
\begin{equation*}
\begin{split}
    \mathrm{RandomForestRegressor}.\mathrm{fit}&(\mathrm{CAGR}(y_1,y_1+\delta),\\
    &[\overline{F}(y_0,y_1), \overline{\Gamma}(y_0,y_1),\\
    &\overline{d}(y_0,y_1), GDPpc(y_1),\\
    &y_0,\ \mathrm{MA}_{fe},\ \mathrm{region}_{fe},\ \mathrm{country_{fe}}]).
\end{split}
\end{equation*}
Because we are interested not only in comparing different features in model training but also in how they impact the output, we make use of SHapley Additive exPlanations (SHAP) analysis \cite{lundberg2017unified}. SHAP is a groundbreaking approach to machine learning interpretability, designed to explain the output of any machine learning model. It is based on the concept of Shapley values, a method from cooperative game theory \cite{shapley1953value} that assigns a fair distribution of both gains and costs to each participant in a coalition. In the context of machine learning, SHAP values measure the impact of each feature in a prediction model. Each feature value of a data instance contributes either positively or negatively to the prediction, compared to the average prediction across the dataset. SHAP effectively decomposes a prediction to show the contribution of each feature to the overall output. One of the key advantages of SHAP is its consistency and local accuracy, ensuring that each feature’s contribution is consistently calculated across different predictions. This makes SHAP particularly useful for providing insights into complex models, such as random forests \cite{deb2021application, hatami2023non, wang2024feature} or neural networks \cite{younisse2022explaining, zheng2022shap, chen2021sales}, where traditional feature importance measures might fall short.

In our work, we use SHAP analysis to understand how Fitness and Coherence affect the CAGR measure. We used the implementation of SHAP in the SHAP Python package (\url{https://shap.readthedocs.io/en/latest/#}) to compute the SHAP values for all features by the command
\begin{equation*}
    \text{shap.TreeExplainer(RandomForestRegressor.fit())}
\end{equation*}
By doing this, we obtain for all three models (one for each geographical scale) the SHAP values for each feature relative to each sample. We do this computation considering all the $y_0$ and $y_1$ with $\delta = 5$ both for MA, region and country. 

First, we show the feature importance comparison in Fig. \ref{fig:RF_FI}. Each pair of bars in this figure represents the feature importance of Fitness, Coherence, diversification, starting GDPpc and geographic and time information, computed with SHAP. The $y$-axis represents the average of the absolute SHAP values for each feature. We observe how the GDPpc($y_1$) is the more important feature, as expected, followed by the year and country geographic information (country$_{\textbf{fe}}$). However, we used them together only to control Fitness and Coherence behaviour. A comparison of only these two quantities is displayed in the main figure of Fig. \ref{fig:RF_FI}. We find that Fitness and Coherence behaviour switch from MA to the country in explaining the trends in CAGR. However, we do not yet know whether this importance translates into a positive or negative effect.\\
\begin{figure*}
    \centering
    \includegraphics[scale=.48]{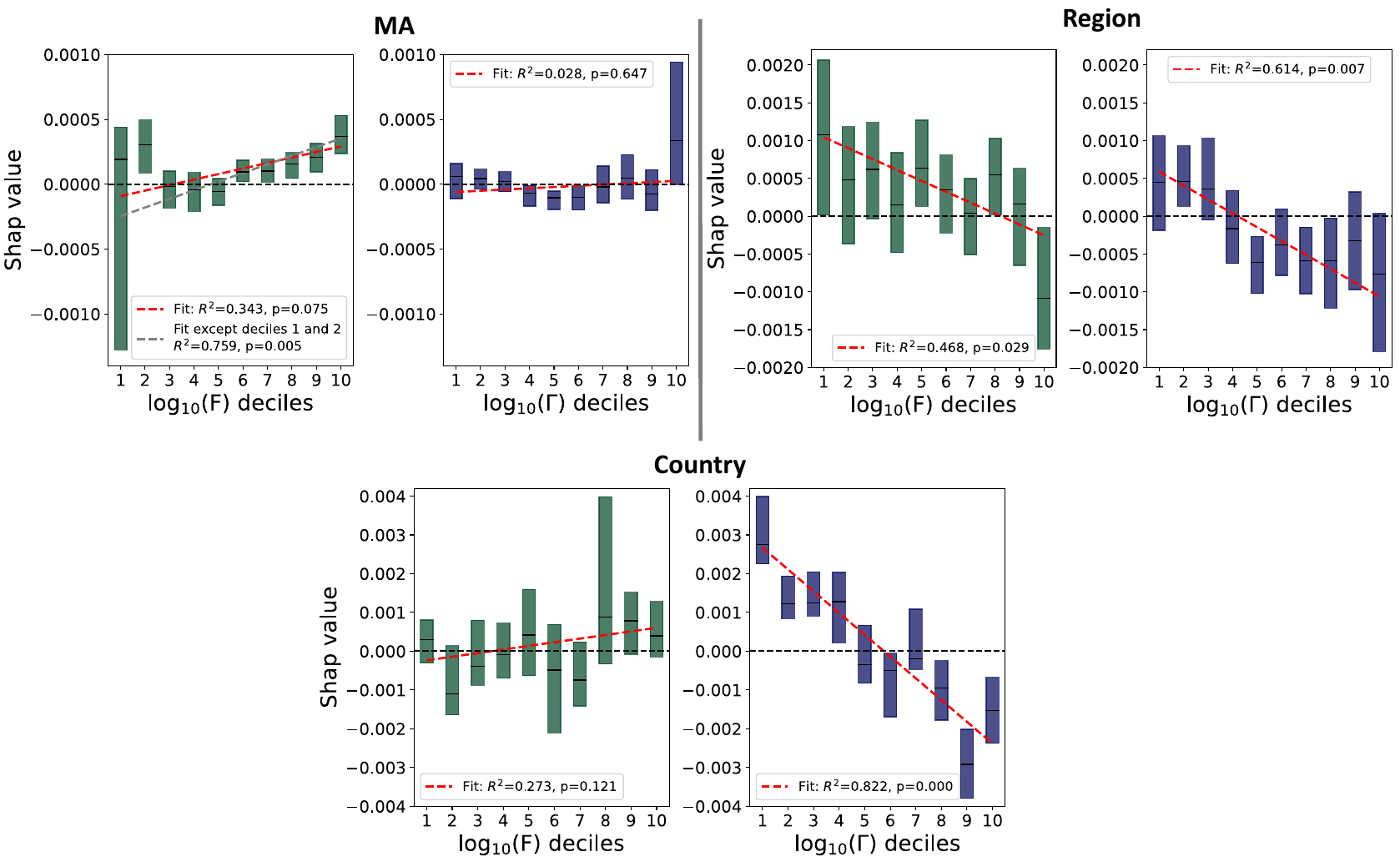}
\caption{\textbf{SHAP values VS $F$ and $\Gamma$ feature value.}
We plot the trends of SHAP values for the Fitness (\(F\)) and Coherence (\(\Gamma\)) features against their respective values across three geographical scales. Each bar in the three couples of figures represents a decile of the dataset, ensuring an equal number of samples per bin. Values below/above zero (marked by the black dotted line) negatively/positively impact the model.  
In addition, we include a weighted linear fit in each plot, reporting the corresponding \(R^2\) and p-value.  
\textbf{MA case}: We performed an additional fit by removing the first two deciles to better highlight the increasing trend and the growing importance of Fitness. While the \(\Gamma\) trend is not statistically significant, we note how high Coherence has a strong positive impact on the economic growth of MAs.
\textbf{Region case}: Both features exhibit a statistically significant downward trend. Notably, the last decile in the \(F\) plot has a substantial negative impact on economic growth.
\textbf{Country case}: $\Gamma$ displays a strong and statistically significant negative trend, underscoring the importance of economic diversification rather than specialization in a narrow range of activities. Conversely, \(F\) shows an increasing but statistically not significant trend. However, it is important to note that our analysis focuses exclusively on highly developed countries. This limitation may reduce the observed role of Fitness in driving economic growth at the national level, as also highlighted in the literature \cite{cristelli2017predictability}.
Another key aspect of these findings that we aim to emphasize is that these measures are highly heterogeneous. Because of this, regressions are not the most appropriate strategy for dealing with these kinds of scenarios \cite{cristelli2015heterogeneous}.
}
\label{fig:SHAP}
\end{figure*}
Finally, we plot in Fig. \ref{fig:SHAP} the trend of the SHAP values of the Fitness and Coherence features against their values, for all three scales. Values less than $0$ impact negatively on the model, positively otherwise. Each bar in all three figures represents a decile of the data, ensuring an equal number of samples per bin.
Furthermore, we apply a weighted linear fit to each plot and report the corresponding \(R^2\) and p-value.  
\begin{itemize}
    \item \textbf{MA case}: We adding a second fit after removing the first two deciles to better emphasize the increasing trend and the growing relevance of Fitness. Although the \(\Gamma\) trend is not statistically significant, we observe that high Coherence has a strong positive effect on the economic growth of MAs.  
    \item \textbf{Region case}: Both features exhibit a statistically significant downward trend. In particular, the last decile in the \(F\) plot has a pronounced negative impact on economic growth.  
    \item \textbf{Country case}: \(\Gamma\) shows a strong and statistically significant negative trend, reinforcing the idea that economic diversification is more beneficial than specialization in a limited set of activities at this level. Conversely, \(F\) displays an increasing trend, although not statistically significant. It is important to emphasize that our analysis focuses exclusively on highly developed countries, which may attenuate the observed role of Fitness in driving national economic growth, as previously noted in the literature \cite{cristelli2017predictability}. 
\end{itemize}
To conclude, it is crucial to emphasize also the high heterogeneity of these measures. As a result, regression techniques may not be the most suitable approach for addressing such scenarios, as discussed in the work of Cristelli et al. \cite{cristelli2015heterogeneous}.

\section{Discussion}
\label{sec:discussion}
The present study examines the impact of two technological measures—Technological Fitness (\(F\)) and Technological Coherence (\(\Gamma\))—on economic growth across different geographical scales (metropolitan areas, regions, and countries). We assess their distinct contributions to growth, as measured by the Compound Annual Growth Rate (CAGR), and elucidate the interplay between technology composition and economic development.
The findings of this study reveal intriguing patterns in the significance of Technological Fitness and Coherence across varying scales.\\
First of all, we make a pairwise correlation analysis between Fitness, Coherence and CAGR as shown in Fig. \ref{fig:F_C_CAGR}. This analysis offers a first qualitative perspective on these interactions, revealing complex and heterogeneous correlation patterns. In particular, we highlight two different trends in MA and region scale. These can suggest two different types of economic development that could be modelled theoretically in another study.\\
Subsequently, the SHapley Additive exPlanations (SHAP) feature importance analysis shows a switched importance between Technological Fitness and Coherence in the correlation with CAGR. However, this result does not clarify whether this importance has a positive or negative impact.
To understand this, we explore how the Fitness and Coherence features influenced growth. 
We find that high Coherence implies faster growth only at the MA scale and this result aligns with previous research emphasizing the importance of Technological Coherence in calculating GDPpc for MAs \cite{straccamore2023urban}. At the regional and national level, Coherence is systematically correlated with lower growth.
The results for Fitness are less sharp: it presents a U-shaped impact on growth for MAs, a decreasing impact on regions, and an upward trend at the national level.


The findings of this study contribute to our understanding of the intricate relationship between technology composition and economic development at different geographical levels. The results imply that the factors driving economic prosperity in metropolitan areas differ from those at regional and national scales.
This heterogeneity of effects implies that policymakers and managers should carefully consider how policies to support innovation and economic development can be improved by taking into account the specificities of Fitness and Coherence.

However, it is important to acknowledge some limitations of this study. The analysis focused solely on the relationship between Technological Fitness, Coherence, and GDPpc, without considering other potentially influential factors such as institutional frameworks, human capital, or infrastructure. Future research could explore the interaction of these additional variables to gain a more comprehensive understanding of the factors driving economic development. Moreover, the study utilized patent data as a proxy for technological capabilities and innovation. While patents provide valuable insights into technological advancements, they may not capture the entire spectrum of technological complexity and innovation. Incorporating other indicators, such as research and development expenditure or technological collaborations, could enhance the accuracy and robustness of the analysis.

In conclusion, this study highlights the varying roles played by Technological Fitness and Coherence in explaining GDPpc across different geographic scales. The findings emphasize the importance of both technological concentration and diversification in driving economic growth, depending on the level of analysis. By understanding these aspects, policymakers and stakeholders can formulate targeted strategies to foster technological progress and regional development. Future research could build upon these findings by incorporating additional factors and exploring different dimensions of Economic Complexity to provide a more comprehensive understanding of the dynamics between technology and economic development.

\bibliography{sample.bib}

\end{document}


\title{\huge{Supplementary Information}\\\Large Comparative Analysis of Technology Fitness and Coherence at different geographical scales}

\author{Matteo Straccamore, Matteo Bruno, and Andrea Tacchella}

\date{}
\maketitle

\newpage

\tableofcontents
\newpage

\newpage

\section{Patents information}

Using the database provided by De Rassenfosse et al.~\cite{de2019geocoding}, we can determine the geographical locations of approximately 18.9 million patent applicants. The objective of this study is to create a dataset comprising early patent applications from various parts of the world and organize it based on the applicants' locations. This dataset enables us to identify the origin of patented inventions and locate centers of innovation. Such information proves valuable for analyzing the geography of innovation and understanding the spatial distribution of patented inventions. Policymakers interested in firms' location decisions and the migration patterns of highly skilled workers can also benefit from this data. The authors accomplished geolocation by associating postal codes of applicant addresses with latitude and longitude coordinates, thereby determining countries, regions, and cities/metropolitan areas. They obtained postal code information by matching addresses found in patent applications within the PATSTAT database and other sources, which provide the corresponding postal codes based on these address codes.\\
Here are some notable features of this database:
\begin{itemize}
\item The patents included in this dataset represent the initial applications for respective inventions.
\item Some patents have multiple geolocations due to collaborative efforts involving multiple applicants. However, it should be noted that after assigning patents to their respective metropolitan areas, only around 2\% of the total patents have multiple geolocations.
\item Each patent is associated with one or more CPC 4-digit technology codes. In cases where patents have multiple technology codes, we retain all of them. This approach adds minimal noise to the data since the categories are already highly specific.
\item As this database focuses on the first application, the applicant's position remains consistent even if multiple patent offices are involved.
\item The geolocalization of all patents in the De Rassenfosse database utilizes information from various sources, including PATSTAT, WIPO, REGPAT, and patent offices in Japan, China, Germany, France, and the United Kingdom.
\end{itemize}

\section{Binarization process}
\begin{figure*}
    \centering
    \includegraphics[scale=.55]{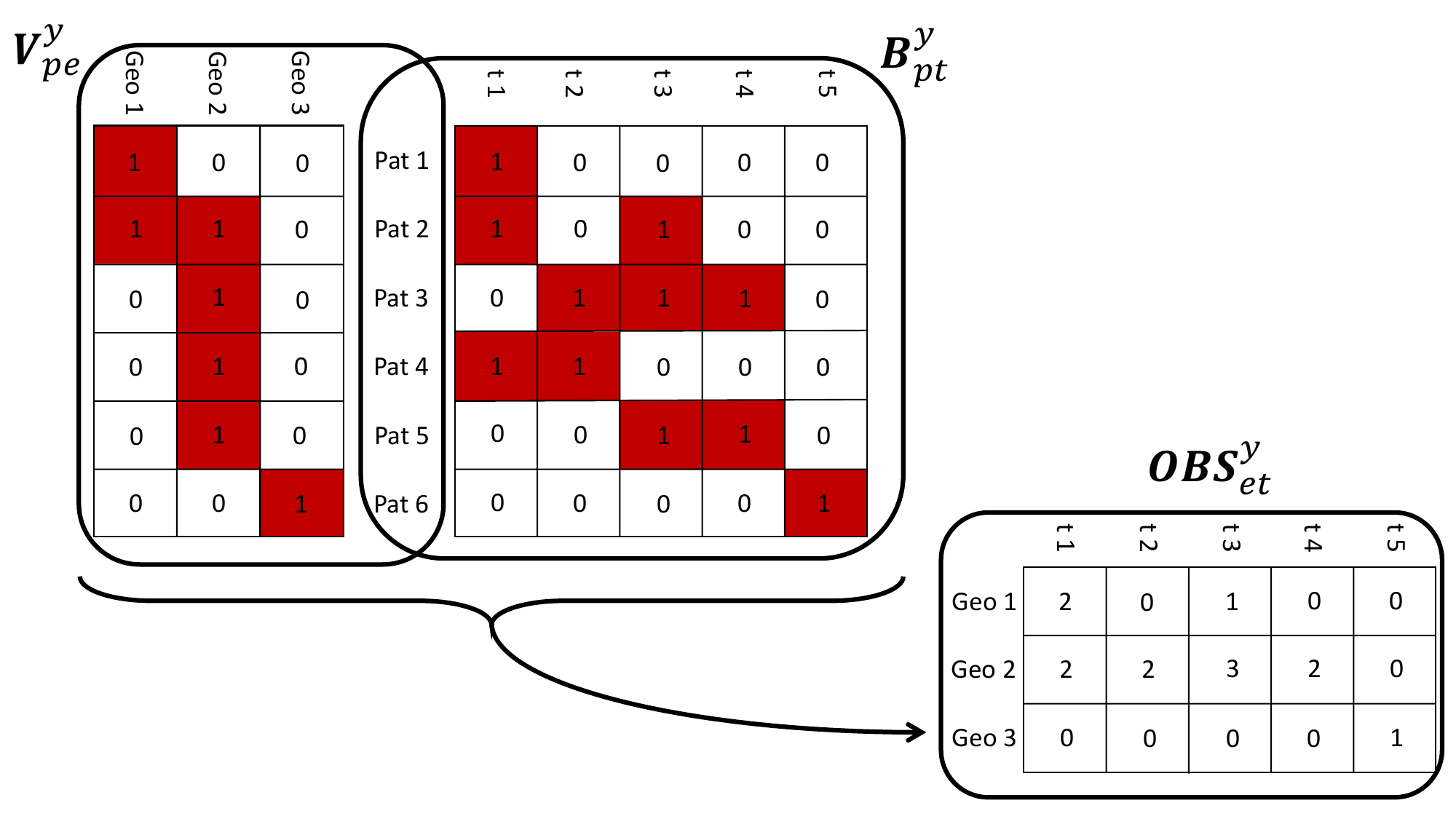}
\caption{\textbf{Observation matrix (OBS) construction process.} In This Figure, we show how we compute the $\textbf{OBS}^y$ matrix for each 5-time window $y$. We start from the bipartites $\textbf{V}^y$ and $\textbf{B}^y$, which elements $V^y_{p,e}$ and $B^y_{p,t}$ are $1$ if an entity $e$, or patent $p$, is connected to a patent $p$, or technology code $t$, respectively. Finally, $\textbf{OBS}^y$ is computed counting the common elements: $OBS^y_{e,t} = \sum_p{V_{p,e}^y B_{p,t}^y}$.}
\label{fig:pval}
\end{figure*}
As described in the main text, the first objective is to generate the bipartite $\textbf{M}^y$ that links the three different geographical areas to their respective technologies. In this work, we decided not to work by calculating the Revealed Comparative Advantage (RCA) \cite{balassa1965trade}. Although the calculation of the RCA has proven to be excellent for studies concerning the export of products~\cite{balassa1965trade, hidalgo2007product, zaccaria2014taxonomy, pugliese2019unfolding}, in our work its application led us to questionable results, such as the United States in the bottom of the Fitness ranking, or the Tokyo metropolitan area in the middle of the respective ranking. The problems concerning the RCA have already been discussed in \cite{bruno2023inferring}. In our case, the problem arises when it is calculated by comparing counts, which range from very low values, such as $1$, to values of very high orders of magnitude, such as $10000$; problem not present in the case of exports at country level.

\begin{figure*}
    \centering
    \includegraphics[scale=.55]{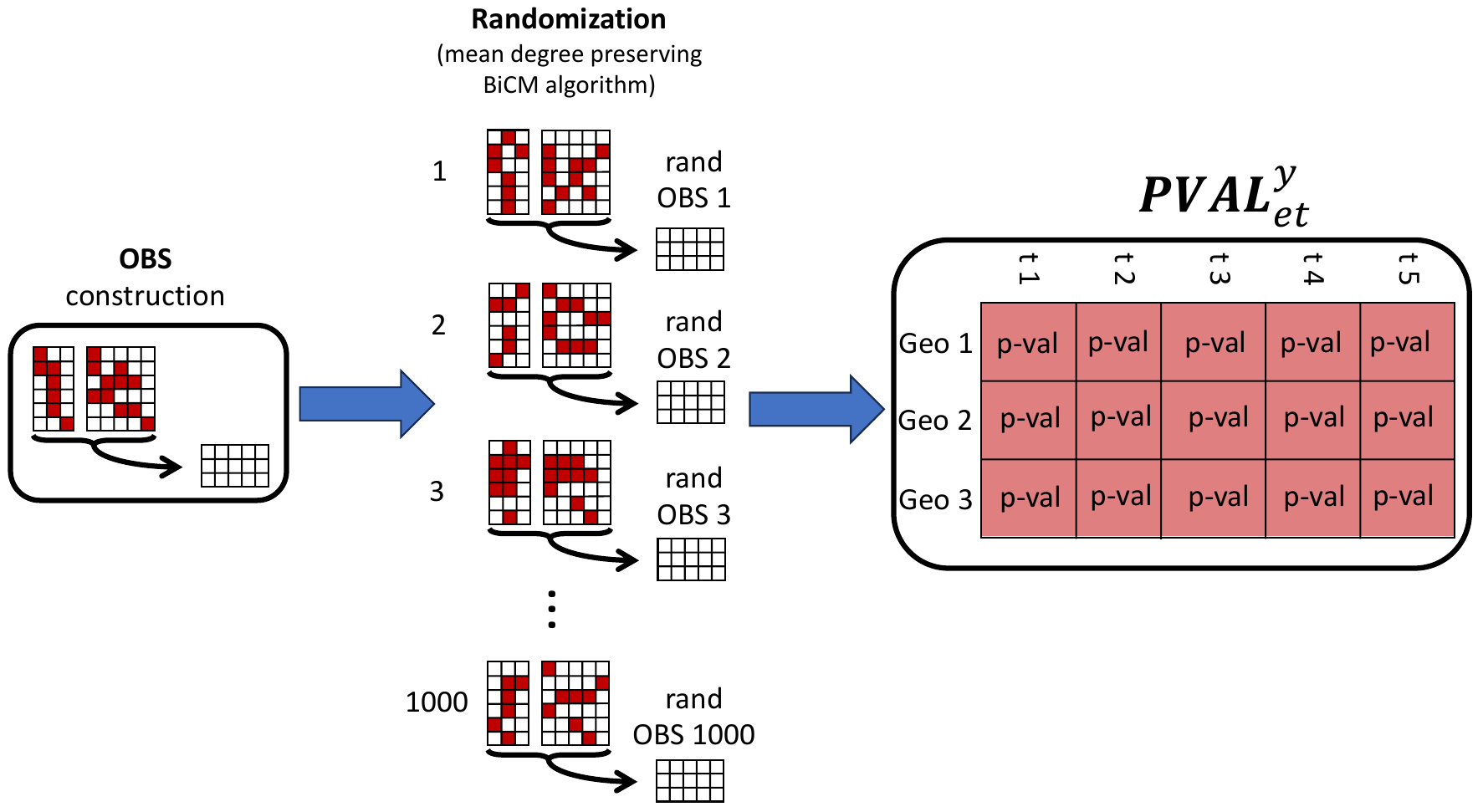}
\caption{\textbf{Statistical validation process of $\textbf{OBS}^y$ matrix.} After the $\textbf{OBS}^y$ computation, we randomize $1000$ times both $\textbf{V}^y$ and $\textbf{B}^y$ by the use of the BiCM algorithm which is used to keep the average grade of the two layers fixed. From these $1000$ matrix pairs, we calculate the respective random observations and conclude by calculating the p-value of the real observations by counting how many times these are greater than the respective random observations. The final output are the $\textbf{PVAL}^y$ matrices, which element $PVAL^y_{e,t}$ is the p-value of the $OBS^y_{e,t}$ observation.}
\label{fig:pval2}
\end{figure*}
To avoid this, in this paper we construct the $\textbf{M}^y$-matrices using an innovative methodology involving statistical tests to validate the number of technologies present in an entity's patents. We start with two preliminary matrices: $\textbf{V}^y$, linking entities $e$ to their patents $p$, and $\textbf{B}^y$, linking patents to their corresponding technologies $t$. In these matrices, $V_{p,e} = 1$ (or $0$) indicates whether patent $p$ belongs to entity $e$, and $B_{p,t} = 1$ (or $0$) if patent $p$ includes technology code $t$.
First, we count the number of observations, i.e. the number of times the technology $t$ falls into the entity $e$. In Fig. \ref{fig:pval} we show a pictorial example of the observation matrix $\textbf{OBS}^y$ construction. We simply compute
\begin{equation*}
    OBS^y_{e,t} = \sum_p{V_{p,e}^y B_{p,t}^y}
\end{equation*}
for each 5-year time window $y$.

After this first process, we validate $\textbf{OBS}^y$ matrix by comparing them against an ensemble of $1000$ matrices generated using the Bipartite Configuration Model algorithm (BiCM), thus obtaining a matrix of p-values, $\textbf{PVAL}^y$. Each element, $PVAL_{e,t}^y$, represents the p-value associated with the link between entity $e$ and technology $t$ for year $y$. This process is show in Fig. \ref{fig:pval2}. The final output is the $\textbf{PVAL}^y$ matrix, which element $PVAL^y_{e,t}$ is the p-value of the $OBS^y_{e,t}$ observation.

We conclude by binarizing $\textbf{PVAL}^y$ using a p-value threshold of $0.05$ as follows:
\begin{equation*}
    M^y_{e,t} = 
    \begin{cases}
        1\ \text{if}\ \text{PVAL}^y_{e,t} \le 0.05\\
        0\ \text{if}\ \text{PVAL}^y_{e,t} > 0.05
    \end{cases}
\end{equation*}
and finally obtaining our $\textbf{M}^y$ matrices.

\section{OLS regression comparison}

\begin{table}[h!]
\centering
\caption{OLS Regression Results - Metropolitan Areas (MA) with year, MA, region, and country as fixed effects}
\begin{tabular}{lcccccc}
\hline
\multirow{2}{*}{Coefficient} & \multicolumn{3}{c}{With Fitness (F) \& Coherence ($\Gamma$)} & \multicolumn{3}{c}{Without Fitness (F) \& Coherence ($\Gamma$)} \\
\cline{2-7}
                             & Estimate & Std. Error & & Estimate & Std. Error & \\
\hline
$F$                            & -0.0337***  & 0.000      & & --       & --         & \\
$\Gamma$                        & 0.0234***   & 0.002      & & --       & --         & \\
div                          & 0.0401***   & 0.001      & & -0.0110***  & 0.000      & \\
GDPpc                        & -0.0138***  & 0.000      & & 0.0061***   & 0.000      & \\
\hline
\multicolumn{4}{l}{Adjusted \( R^2 \): 0.708} & \multicolumn{3}{l}{0.464} \\
\hline
\multicolumn{7}{l}{\footnotesize *** $p<0.001$, ** $p<0.01$, * $p<0.05$} \\
\end{tabular}
\end{table}

\begin{table}[h!]
\centering
\caption{OLS Regression Results - Regions with year, region, and country as fixed effects}
\begin{tabular}{lcccccc}
\hline
\multirow{2}{*}{Coefficient} & \multicolumn{3}{c}{With Fitness (F) \& Coherence ($\Gamma$)} & \multicolumn{3}{c}{Without Fitness (F) \& Coherence ($\Gamma$)} \\
\cline{2-7}
                             & Estimate & Std. Error & & Estimate & Std. Error & \\
\hline
$F$& -0.0401***  & 0.001      & & --       & --         & \\
$\Gamma$& -0.0013      & 0.005      & & --       & --         & \\
div                          & 0.0488***   & 0.001      & & -0.0032***  & 0.000      & \\
GDPpc                        & -0.0168***  & 0.000      & & 0.0040***   & 0.000      & \\
year                         & -0.0020***  & 0.000      & & -0.0016***  & 0.000      & \\
\hline
\multicolumn{4}{l}{Adjusted \( R^2 \): 0.498} & \multicolumn{3}{l}{0.336} \\
\hline
\multicolumn{7}{l}{\footnotesize *** $p<0.001$, ** $p<0.01$, * $p<0.05$} \\
\end{tabular}
\end{table}

\begin{table}[h!]
\centering
\caption{OLS Regression Results - Countries with year and country as fixed effects}
\begin{tabular}{lcccccc}
\hline
\multirow{2}{*}{Coefficient} & \multicolumn{3}{c}{With Fitness (F) \& Coherence ($\Gamma$)} & \multicolumn{3}{c}{Without Fitness (F) \& Coherence ($\Gamma$)} \\
\cline{2-7}
                             & Estimate & Std. Error & & Estimate & Std. Error & \\
\hline
$F$& -0.0245***  & 0.002      & & --       & --         & \\
$\Gamma$& 0.0121       & 0.029      & & --       & --         & \\
div                          & 0.0248***   & 0.002      & & 0.0041*   & 0.002      & \\
GDPpc                        & -0.0093***  & 0.001      & & 0.0004    & 0.001      & \\
year                         & -0.0019***  & 0.000      & & -0.0013*** & 0.000      & \\
\hline
\multicolumn{4}{l}{Adjusted \( R^2 \): 0.703} & \multicolumn{3}{l}{0.471} \\
\hline
\multicolumn{7}{l}{\footnotesize *** $p<0.001$, ** $p<0.01$, * $p<0.05$} \\
\end{tabular}
\end{table}

\newpage

\bibliography{sample.bib}